%% file: main.tex
\documentclass{article}
\usepackage{threeparttable}
\usepackage{spconf,amsmath,graphicx}
\usepackage{multirow, array}
\usepackage{flushend}
\usepackage{amssymb}
\usepackage{enumitem}
\usepackage{bm} 
\usepackage{cite}
\usepackage{xcolor}
\usepackage{array}
\usepackage{flushend}
\usepackage{algorithm}
\usepackage{algpseudocode}
\usepackage{hyperref}
\usepackage{subcaption}
\usepackage{tabularx} 

\title{Towards Lightweight and Stable Zero-shot TTS with Self-distilled Representation Disentanglement}

\name{Qianniu Chen$^{1,2}$, Xiaoyang Hao$^{2}$, Bowen Li$^{2}$, Yue Liu$^{2}$, Li Lu$^{* 1}$\thanks{$^*$Corresponding author.}\thanks{$^1$The authors are with the State Key Laboratory of Blockchain and Data Security, Zhejiang University \& Hangzhou High-Tech Zone (Binjiang) Institute of Blockchain and Data Security, Hangzhou, China}}

\address{
$^{1}$Zhejiang University, China \qquad
$^{2}$AMAP, China}

\begin{document}
\ninept
\maketitle
 \begin{abstract}
     Zero-shot Text-To-Speech (TTS) synthesis shows great promise for personalized voice customization through voice cloning. 
    However, current methods for achieving zero-shot TTS heavily rely on large model scales and extensive training datasets to ensure satisfactory performance and generalizability across various speakers. This raises concerns regarding both deployment costs and data security.
    In this paper, we present a lightweight and stable zero-shot TTS system. We introduce a novel TTS architecture designed to effectively model linguistic content and various speaker attributes from source speech and prompt speech, respectively. Furthermore, we present a two-stage self-distillation framework that constructs parallel data pairs for effectively disentangling linguistic content and speakers from the perspective of training data.
    Extensive experiments show that our system exhibits excellent performance and superior stability on the zero-shot TTS tasks. Moreover, it shows markedly superior computational efficiency, with RTFs of 0.13 and 0.012 on the CPU and GPU, respectively.

\end{abstract}


\input{sections/1_introduction}

\input{sections/2_Design}

\input{sections/3_Evaluation}

\input{sections/4_Conclusion}

\clearpage

\bibliographystyle{IEEEbib}
\bibliography{bib}

\end{document}

%% file: sections/1_introduction.tex
\section{Introduction}
The recent advancements in Text-To-Speech (TTS) synthesis have catalyzed an increasing demand for personalized voice customization, especially in speaker imitation using voice cloning techniques. 
However, traditional TTS approaches typically require a large amount of high-quality speaker speech data as supervision signals to adapt to new speakers\cite{Cong2020,Ren2021}. Such approaches not only demand substantial resources to fine-tune and maintain a unique TTS model for each user, but also require users to record substantial amounts of speech data.
To alleviate this challenge, recent research\cite{Casanova2021,Wu2022,Casanova2022,Li2023, Ju2024, Jiang2024, 2022Huang, lee2023, 2023Kang, Shen2024, 2024Lee,Wang2023,Chen2024,CosyVoice,GPT-SoVITS,seedtts} has explored zero-shot voice cloning with multi-speaker TTS systems, appearing to provide a promising solution. 

Zero-shot TTS provides a highly generalised model that is capable of synthesising speech that resemble a new speaker with only a brief prompt speech (typically a few seconds). This approach eliminates the need for additional model finetuning.
Recent zero-shot TTS studies can be broadly categorized into two types: those leveraging speaker representation models\cite{Ju2024,Casanova2021,Wu2022,Casanova2022,Li2023, Jiang2024} and those employing in-context learning strategies\cite{Wang2023,Chen2024,CosyVoice,GPT-SoVITS,seedtts}. Speaker representation-based approaches enable explicitly modeling a target speaker's representation from any given speech input. This is achieved either through pre-trained speaker encoders\cite{Jia2018, Arik2018, Casanova2022, Kumar2022} or through speaker encoders jointly trained with the TTS model\cite{2022Huang, lee2023, Jiang2024, 2023Kang, Li2023}. These speaker representations are then utilized as global\cite{Casanova2022, Wu2022} or temporal\cite{Jiang2024, Shen2024} conditional encodings within the audio generation. By thoroughly learning the speaker characteristics of numerous speakers in the training dataset, the model can map the speaker latent space to the corresponding speech properties, thus attaining zero-shot capability. 
Recent studies leverage superior in-context learning capabilities of diffusion models\cite{Ju2024, 2024Lee} or Large Language Models (LLMs)\cite{Wang2023,Chen2024,CosyVoice,GPT-SoVITS,seedtts} to learn speaker-related characteristics from the prompt speech. These approaches demonstrate the efficacy of leveraging contextually rich information to achieve speech synthesis across unseen speakers.

Despite significant progress in recent zero-shot TTS, these methods still face considerable challenges in practical applications, i.e., high resource dependence and suboptimal synthesis stability. 
First, current zero-shot TTS systems rely on large-scale parameters to model the inherent diversity of human voices from a massive amount of data.
For example, CosyVoice\cite{CosyVoice} employs a 414M model and requires 171K hours of training data. Such an enormous model hinders their integration into resource-constrained environments. It not only increases service costs,  but also raises privacy and security concerns, as users' speech prompts still need to be uploaded to cloud servers.
Second, while autoregressive speech modeling methods such as Tacotron\cite{Wang17} and VALL-E\cite{Wang2023} improve in-contextual modeling and enhance speech expressiveness, they also increase vulnerability to time series prediction errors such as omissions, incorrect readings, repetitions, and even severe error accumulation, leading to suboptimal stability.

In this paper, we present a lightweight and stable zero-shot TTS system with self-distilled representation disentanglement.
On the one hand, we propose a TTS architecture that effectively models the prompt speech into multi-level speaker representations, including global timbre features and temporal style features (i.e., features that change over time, including speed, pitch, and energy). 
By explicitly leveraging the multi-level speaker representations, the model can operate effectively with less data and simpler architectures, as these representations capture the essential characteristics of the speakers.
On the other hand, we employ a self-distillation framework to enhance the model's ability to disentangle content from speaker characteristics. We use a pre-trained teacher model to generate parallel data pairs that differ only in speaker characteristics, thereby guiding the training of a student model. This disentangles the linguistic content from the speaker-related attributes at the data level, reinforcing the model's ability to synthesize speech conditioned solely on the provided prompt speech. 
Extensive objective and subjective evaluations demonstrate that our system outperforms baseline models in content integrity while showing promise in speaker similarity. Moreover, our framework exhibits the most exceptional computational efficiency, rendering it well-suited for resource-constrained environments and real-time applications.
Audio samples are available at the demo website: \url{https://zzfng9696.github.io}.

%% file: sections/2_Design.tex
\begin{figure}[t]
    \centering
    \includegraphics[width=\linewidth]{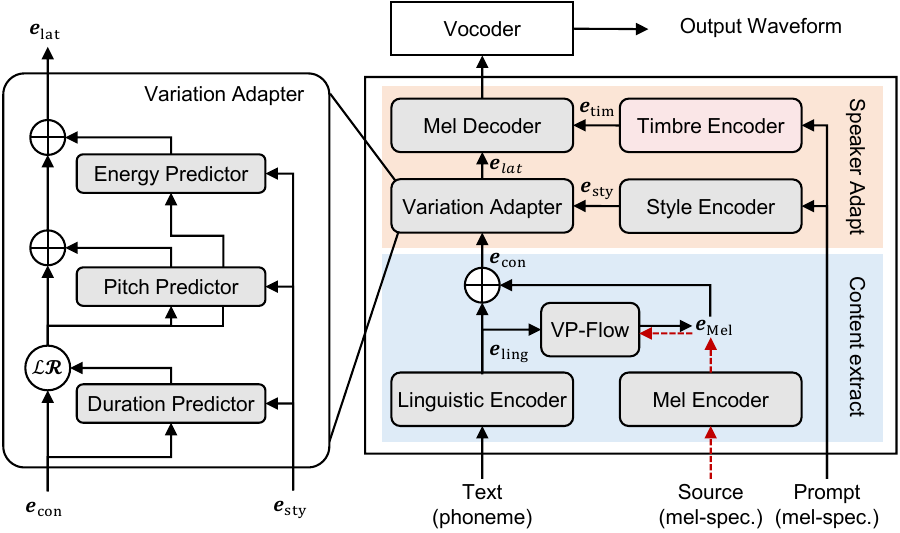}
    \caption{Architecture of the proposed TTS model. The red dashed line indicates participation only during training. The Timbre encoder is a pre-trained model requiring no optimization.}
    \label{fig:arch}
\end{figure}
\begin{figure*}[t]
    \centering
    \includegraphics[width=\linewidth]{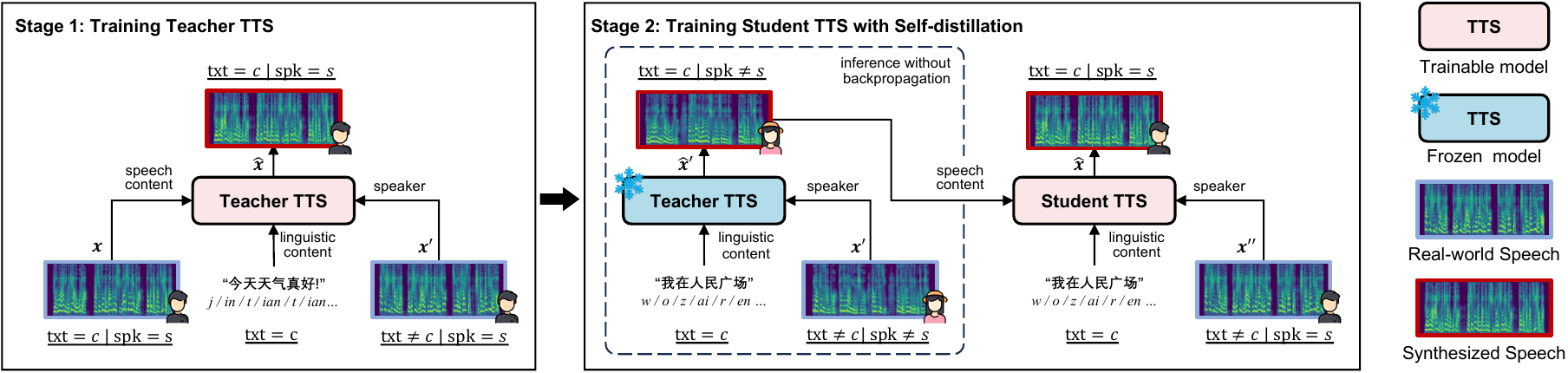}
    \caption{Two-stage training framework.}
    \label{fig:overall}
    \vspace{-3mm}
\end{figure*}

\section{Zeo-shot TTS Model}

In this section, we introduce the model of our system. Figure \ref{fig:arch} shows the model architecture, which comprises two main parts, i.e., content extraction and speaker adaptation.

\subsection{Content Extraction}
In the first part, we extract speaker-independent content representation from both text (phoneme sequence) and the corresponding ground truth speech (mel-spectrogram).
To enhance the expressive capacity for complex speech content, we integrate a Mel Variational Autoencoder (VAE) with a flow-based model to construct a latent space of speech, transcending the limitations of text-only representations. 
We first encode both the text and speech at the phoneme level using a linguistic encoder and a mel encoder, generating representations $ \bm{e}_{\text{ling}} $ and $ \bm{e}_{\text{mel}}$, respectively.
We use $ \bm{e}_{\text{ling}} $ as a condition for predicting $ \bm{e}_{\text{mel}} $ through Volume-Preserving Flow (VP-Flow). After that, we fuse $ \bm{e}_{\text{ling}}$ with $ \bm{e}_{\text{mel}} $, forming the content representation $ \bm{e}_{\text{con}} $, which is then fed into the speaker adaptivation part. Specifically, the text encoder consists of multiple feed-forward Transformer blocks with relative positional encoding\cite{Shaw18}.
The mel encoder consists of multiple 2D Residential blocks (ResBlock) to downsample and encode the input mel-spectrogram, followed by average pooling to map frame-level features to phoneme-level features. Another multi-level ResNet estimates the mean and variance of the variational latent space. Note that the source speech contains not only content information, but also information related to the target speaker. This content-speaker coupling leads to the leakage of speaker information from $ \bm{e}_{\text{con}} $. We address this issue using a self-distillation framework in Section \ref{sec:dist}.

\subsection{Speaker Adaptation}
In the second part, we explicitly model multi-level speaker-specific representations from the prompt speech, i.e., temporal style representations and global timbre representations. These representations serve as conditions for steering the synthesis of the target speech, ensuring that it accurately reflects the desired speaker characteristics.

We adopt a trainable mel encoder to extract the temporally-related style representation, $\bm{e}_{\text{sty}}$. This encoder is similar to that used in HierSpeech++\cite{lee2023} and GPT-SoVITS\cite{GPT-SoVITS}, but it includes positional encoding and omits the final temporal averaging layer to better capture temporal information. 
We use $\bm{e}_{\text{sty}}$ as a condition for a variation adapter, which consists of three predictors for independent speech attribute predictors, i.e., duration, pitch, and energy. We integrate and align $\bm{e}_{\text{sty}}$ with the input sequence of the predictors through a multi-head cross-attention layer. These predictors are individually trained under supervised conditions using annotated data. To ensure the model's stability, we separate the gradients of these predictors from the main network.

For timbre extraction, we utilize a speaker recognition model as the timbre encoder, i.e., ERes2NetV2\cite{chen2024eres2netv2}, pre-trained on a large-scale multi-domain dataset, Speaker3D\cite{zheng20233d}. This encoder generates a comprehensive speaker latent space that encompasses various speakers and speaking conditions, facilitating the effective application of unseen speaker features to achieve precise modeling of their vocal styles. Additionally, a trainable linear layer is integrated into this model to yield the speaker representation $\bm{e}_{\text{spk}}$.
To transfer the timbre representation $\bm{e}_{spk}$ extracted from arbitrary speech into synthesized speech, we employ a speaker-adaptive mel Decoder based on a multi-level ResNet architecture with Adaptive Instance Normalization (AdaIN) layer.

\section{Two-stage Self-distillation}
\label{sec:dist}
In this section, we propose a self-distillation framework aimed at disentangling linguistic content and speaker characteristics from the perspective of data.
The basic idea is to construct parallel data pairs featuring identical content but distinct speakers, utilizing the model's initial zero-shot capability. 
Figure~\ref{fig:overall} shows the training framework, which involves the sequential training of a teacher model and a student model. 

\subsection{Stage 1: Training Teacher TTS} \label{train_tts}

We begin by training a teacher model $\text{M}_T$, which is expected to effectively reconstruct the linguistic content from the given text and preliminarily clone the speaker's voice from the prompt speech.

Given a single speech sample $ x \in \mathbb{R}^{T \times F} $, where $ T $ and $ F $ represent the temporal and frequency dimensions, respectively, we extract $\bm{e}_\text{mel}$ from $ x $ and $\bm{e}_\text{ling}$ from the corresponding text $ c $. These two representations are combined to obtain $\bm{e}_\text{con}$. Next, we use the timbre encoder and the style encoder to extract $\bm{e}_\text{tim}$ and $\bm{e}_\text{sty}$ from the prompt speech $ x' $. Using $\bm{e}_\text{sty}$ as a condition, we predict time-dependent speech attributes and apply these to $\bm{e}_\text{con}$. Finally, we input the fused feature into the Mel decoder to reconstruct the speech mel spectrogram, ensuring that the generated speech matches the target speaker's timbre by conditioning on $\bm{e}_\text{tim}$. We employ the following loss functions to guide the training process.

\textbf{Primary Losses}.
A reconstructive loss ($L_{\text{rec}}$) based on the $L_1$-norm is employed to quantify the discrepancy between the reconstructed and original speech. 
Additionally, a Kullback-Leibler (KL) divergence loss ($L_{\text{KL}}$) is utilized to constrain the distribution of the speech and linguistic representations. 
A PatchGAN-structured discriminator is employed to leverage its capability for multi-scale analysis, distinguishing synthesized mel-spectrograms from genuine audio features. This discriminator facilitates an adversarial training paradigm with the generator, utilizing the Least Squares Generative Adversarial Network (LSGAN) loss ($L_{\text{adv}}$).

\textbf{Style-related Losses}.
A suite of loss functions refine the accuracy of our speech attribute predictors within the variance adapter. For every speech sample $x$, our model generates predictions for the attributes, i.e., duration, pitch, and energy, denoted as $\hat{y}_{i}$, where $i$ spans \{duration, pitch, energy\}. The loss function, $L_{\text{pred},i}$, quantifies the mean squared error (MSE) between the model's predictions and the ground truth values.

\textbf{Timbre-related Loss}.
A cyclic contrastive loss function $L_{\text{cyc}}$ is employed to enforce alignment between the output speech and the prompt audio within the speaker space of the timbre encoder. Let $\bm{e}_i$ and $\bm{\hat{e}}_i$ denote the timbre representations of the original mel-spectrogram and reconstructed mel-spectrogram, respectively, where $i$ indexes the samples within a batch of size $B$. The speaker representation $\bm{\hat{e}}_i$, obtained by feeding $\hat{x}_i$ into the speaker encoder, should closely match $\bm{e}_i$. Conversely, speaker representations of other utterances $\bm{\hat{e}}_{j\neq i}$ within the mini-batch should significantly diverge. The contrastive loss is defined as\cite{Chen2020}:
\begin{equation}
L_{\text{cyc}} = -\frac{1}{B} \sum_{i=1}^{B} \log \left( \frac{\exp(\cos(\bm{\hat{e}}_i, \bm{e}_i))}{\sum_{j=1}^N \mathbf{1}_{j \neq i} \exp(\cos(\bm{\hat{e}}_i, \bm{e}_j))} \right),
\end{equation}
where, $\cos(\cdot, \cdot)$ denotes the cosine similarity between two vectors, and $\mathbf{1}_{j \neq i} \in \{0, 1\}$ is an indicator function. 




\subsection{Stage 2: Training Student TTS}

By leveraging the well-trained teacher model, we proceed to generate parallel data pairs for distilling a student model.

Given a speech sample $x$, and the corresponding text, represented as a phoneme sequence $c$, our method begins by employing the teacher model to synthesize a new speech sample $\hat{x}'$. The synthetic speech sample preserves the textual content $c$, but is delivered by a distinct speaker.
To this end, we begin by randomly selecting a prompt speech sample $x'$ from our training dataset. The sample $x'$ corresponds to a different speaker $s'$ than the speaker $s$ in $x$. Then, we infer the teacher model with $c$ and $x'$ for the generation of $\hat{x}' = \text{M}_T(c, x')$. For the strict temporal alignment of the phonemes of the $x$ and $\hat{x}'$, we explicitly specify the prediction result of the duration predictor as the annotated phoneme duration of $x$ during the inference process.

After obtaining the time-aligned parallel data pair $\{x, \hat{x}'\}$, we proceed to train the student model.
The student model mirrors the same model structure as the teacher model but is initialized randomly. We input $\hat{x}'$ as the speech content for the student model, accompanied by $c$ as the linguistic content input. Additionally, we randomly select another speech sample $\hat{x}''$ as the prompt, which is uttered by the same speaker as $x$.
We optimize the student model following the same procedure as detailed in Section \ref{train_tts}, using $x$ along with its detailed annotation serving as the ground truth.

To regulate the intensity of distillation, we blend real and synthesized samples within each training batch. To this end, we introduce a coefficient $\sigma \in [0, 1]$, which modulates the proportion of synthetic speech samples. Specifically, in each batch of speech content input, we replace $\lfloor\sigma \cdot B\rfloor$ randomly selected speech samples with their corresponding synthesized speech samples, while the remaining samples remain as the ground truth speech samples. 

The self-distillation framework ensures that the student model's content extraction avoids speaker-specific details, compelling it to rely solely on prompt speech for speaker adaptation, thereby enhancing the separation of the representations.

%% file: sections/3_Evaluation.tex
\section{Evaluation}
    \subsection{ Experiemnt Setup}
    We train both the teacher and student models on a diverse dataset, comprising 4,678 speakers, totaling 531 hours of audio.
    The preprocessing pipeline involve extracting 80-dimensional mel-spectrograms from the audio clips. For audio reconstruction from mel-spectrograms, we employ a NFS-HiFiGAN model\cite{wangNSFall} as vocoder. The TTS models are trained using an AdamW optimizer and a batch size of 64 for 200,000 training steps. When training the student model, we set $\sigma=0.8$ by default.

    Our proposed model is compared against four state-of-the-art zero-shot Text-to-Speech (TTS) models: Vall-E\cite{Wang2023}, X-TTSv2\cite{xtts}, CosyVoice\cite{CosyVoice}, and GPT-SoVITS\cite{GPT-SoVITS}. We reproduce Vall-E based on the published methodology\cite{Wang2023} and trained it on a speech corpus of approximately 60K hours and 19K speakers. The other systems utilize the open-source code and pre-trained models provided by their respective authors.

    Prompt speeches for the test dataset are collected from 20 volunteers (10 males and 10 females) who are not part of the training set, ensuring that the baseline system’s training data does not contain these test cases. For each volunteer, we record 5 audio clips, ensuring high audio quality, clear pronunciation, and varied durations ranging from 5 to 10 seconds. For each prompt speech, we generate 100 synthetic speeches using different sentences, with lengths ranging from approximately 30 to 50 characters.

    The metrics employed in the experiments are as follows,
    \begin{itemize}[leftmargin=*]
      \item \textbf{Speaker Similarity (SIM)}: We quantified the similarity between generated and prompt speech using cosine similarity of speaker embeddings, extracted via an Ecapa-TDNN model\cite{ecapaTDNN} pre-trained on Speaker3D dataset\cite{zheng20233d}.
      \item \textbf{Character Error Rate(CER)}: 
        We use a pre-trained ASR model\cite{gao2022paraformer} to infer transcripts, and then derive CERs by calculating the edit distance between the transcripts and the correct text.
      \item \textbf{Real-Time Factor(RTF)}: We measure RTF on a Sever platform equipped with an Intel(R) Xeon(R) Platinum 8369B@2.90GHz CPU and an Nvidia Tesla A100 GPU.
      \item \textbf{Mean Opinion Score (MOS)}: We invite 50 volunteers to score 20 groups of generated speech based on their on their subjective feelings in terms of content consistency ($\text{MOS}_\text{con}$) and speaker similarity ($\text{MOS}_\text{sim}$) on a scale of 1 to 5. Final MOS scores are calculated with a 95\% confidence interval.
    \end{itemize}

\subsection{Overall Performance Comparing with Baselines}

\begin{table}[b]
  {
  
  \centering
  \caption{Performance comparison of our system with baselines}
  \begin{tabularx}{\linewidth}{c|>{\centering\arraybackslash}p{0.7cm}|>{\centering\arraybackslash}X|>{\centering\arraybackslash}p{0.7cm}|>{\centering\arraybackslash}X}
    
    \hline
    \multirow{2}{*}{\textbf{Model}} & \multicolumn{2}{c|}{\textbf{Content Integrity}} & \multicolumn{2}{c}{\textbf{Speaker Similarity}} \\ \cline{2-5}
    & \textbf{CER$\downarrow$} & \textbf{$\text{MOS}_\text{con}\uparrow$} & \textbf{SIM$\uparrow$}  & \textbf{$\text{MOS}_\text{sim}\uparrow$} \\
    \hline
    Vall-E      & 2.89    & 4.17$\pm$0.11     & 0.72                        & 3.24$\pm$0.12  \\
    X-TTSv2     & 4.05    & 3.31$\pm$0.16     & 0.47                        & 2.04$\pm$0.09  \\
    GPT-SoVITS  & 4.89    & 4.18$\pm$0.12     & 0.70                        & 3.40$\pm$0.12  \\
    CosyVoice   & 3.42    & 4.35$\pm$0.10     & \textbf{0.84}               & \textbf{3.73$\pm$0.11}  \\ 
    \hline 
    \textbf{Our system} & \textbf{1.80} & \textbf{4.43$\pm$0.09}  & 0.73    & 3.31$\pm$0.11   \\
    \hline
  \end{tabularx}
  \label{tab:results}
  \vspace{3mm}
  \centering
  \caption{Efficiency comparison of our system with baselines}
\begin{tabular}{c|c|c|c|c} 
    \hline
    \textbf{Model} & \textbf{Params$\downarrow$} & \textbf{Data(h)$\downarrow$} & \textbf{$\text{RTF}_{\text{CPU}}$$\downarrow$} & \textbf{$\text{RTF}_{\text{GPU}}$$\downarrow$} \\
    \hline
    Vall-E      & 370M  & 100K & 2.61 & 0.47 \\
    X-TTSv2     & 447M  & N/A & 5.75 & 0.26 \\
    GPT-SoVITS  & 223M  & 2K & 3.53 & 0.38 \\
    CosyVoice   & 414M  & 173K & 15.3 & 1.89 \\
    \hline
    \textbf{Our system} & \textbf{22.5M} & \textbf{531} & \textbf{0.13} & \textbf{0.012} \\
    \hline
  \end{tabular}
  \label{tab:results2}
  }
\end{table}
Table~\ref{tab:results} presents the overall performance comparison of our system with baseline systems regarding content integrity and speaker similarity. In terms of content integrity, our system achieves the lowest CER of 1.8 and the highest $\text{MOS}_\text{con}$ of 4.43, surpassing all baseline models. These results highlight our system's ability to generate accurate and reliable content that meets expectations. In terms of speaker similarity, our system attains a SIM score of 0.73, trailing only behind the state-of-the-art model, CosyVoice, which scores 0.84. $\text{MOS}_\text{sim}$ may slightly differ from the SIM results, because the objective scoring model primarily assesses timbre similarity, while human raters consider additional factors such as speaking style and emotional expressiveness. 
Overall, the experimental results demonstrate that our system outperforms baseline models in content integrity while showing promise in speaker similarity.

Table~\ref{tab:results2} shows the efficiency comparison of our system with baseline systems in terms of parameter scale, data requirements, and real-time performance metrics. All baseline systems exceed 200M parameters, with CosyVoice and X-TTSv2 even surpassing 400M parameters. While this extensive parameter count enables robust speaker modeling, it necessitates considerable training datasets and computational resources, as illustrated by the significant data hours and RTF involved. In contrast, our system employs a lightweright architecture with only 22.5M parameters, making it both data and computationally efficient. Regarding real-time performance, all baseline systems operate slower than real-time on CPUs, with their RTF values significantly exceeding 1. In comparison, our system achieves RTFs of 0.13 on CPUs and 0.012 on GPUs, indicating at least a tenfold improvement over the baselines. 

\subsection{Impact of Self-distillation Framework}

\begin{figure}[t]
  \centering
  \begin{subfigure}[b]{0.49\linewidth} 
    \centering
    \includegraphics[width=\linewidth]{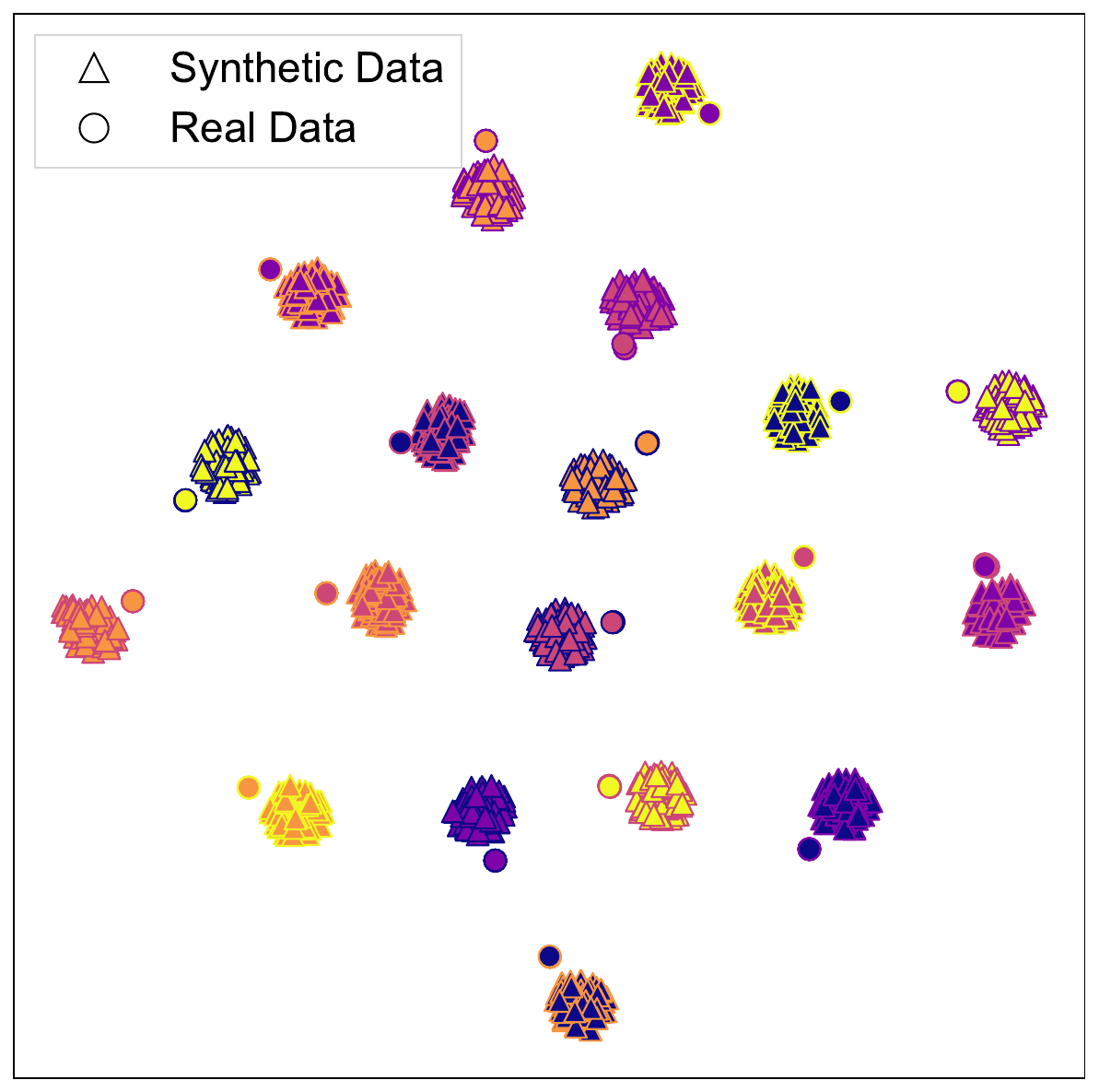}
    \caption{Without self-distillation}
    \label{fig:tsne_with}
  \end{subfigure}
  \hfill 
  \begin{subfigure}[b]{0.49\linewidth}
    \centering
    \includegraphics[width=\linewidth]{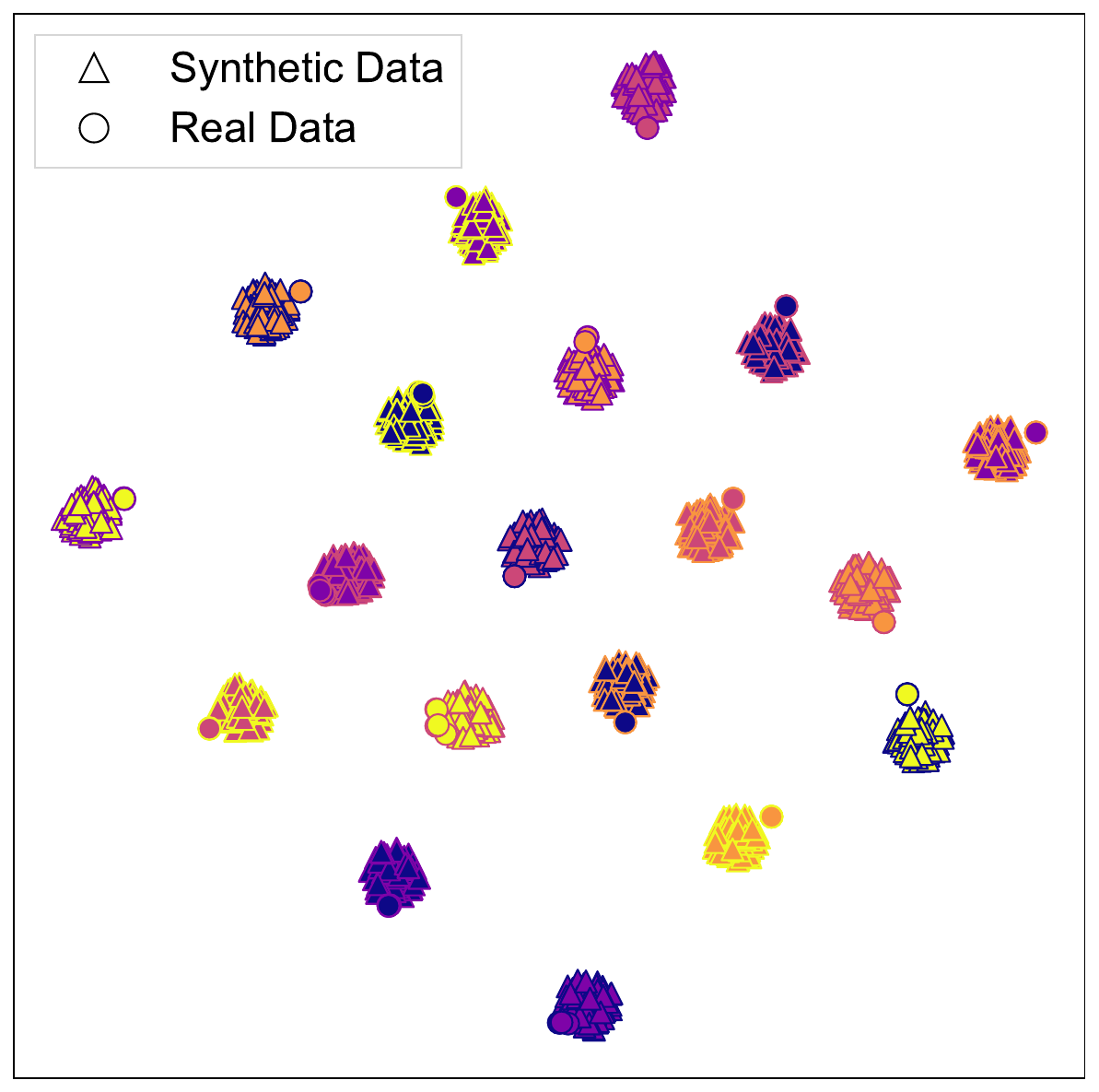} 
    \caption{With self-distillation}
    \label{fig:tsne_without}
  \end{subfigure}
  \vspace{-0mm}
  \caption{t-SNE visualization of speaker embeddings from our system with and without self-distillation. Triangles represent synthetic samples, circles represent real samples.}
  \label{fig:tsne}
  
\end{figure}

\begin{figure}[t]
  \centering
  \includegraphics[width=\linewidth]{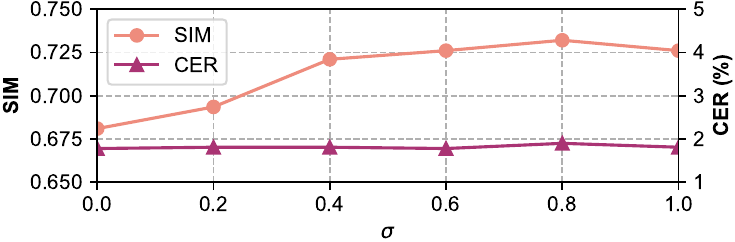}
  \caption{Impact of the self-distillation coefficient $\sigma$ on SIM and CER.}
  \label{fig:sim}
  
\end{figure}

Figure \ref{fig:tsne} shows the t-SNE visualization of speaker embeddings from our system with and without self-distillation. 
We can obvious that,
compared with real speech samples(triangles), the non-distillation model shows a clear separation between the point clusters of real speeches and synthesized speeches (circles). In contrast, the self-distilled model exhibits less separability. The distance between the feature points of real speech and synthesized speech is significantly reduced, and in some cases, they even highly overlap. This indicates a high degree of timbre consistency achieved by the self-distilled model, highlighting its effectiveness in enhancing the similarity of model speaker characteristics.

Figure~\ref{fig:sim} quantitatively illustrates the impact of the self-distillation coefficient $\sigma$ on the SIM and the CER. 
As $\sigma$ increased, the SIM score steadily improves to a maximum of 0.73 at $\sigma=0.8$. Further increases in $\sigma$ led to a slight decline in the score.
This trend emphasizes the importance of selecting an appropriate $\sigma$ value to enhance speaker feature disentanglement and improve the model's generalization capabilities.
Meanwhile, CER scores exhibit stability across varying $\sigma$ values, experiencing minimal fluctuations between $1.78$ and $1.90$. This finding indicates that the self-distillation framework can effectively improve victim similarity without compromising the integrity of the content.

%% file: sections/4_Conclusion.tex
\section{Conclusion}
In this paper, we introduce a lightweight and stable zero-shot TTS system. We first propose a novel TTS architecture designed to effectively model linguistic content and various speaker-related attributes from source speech and prompt speech, respectively. Next, we present a two-stage self-distillation framework that constructs parallel data pairs for effectively disentangling linguistic content and speakers from the aspect of training data. Both objective and subjective evaluations demonstrate that our system matches the Baseline in zero-shot speech synthesis performance while achieving superior computational efficiency.

%% file: main.bbl
\begin{thebibliography}{10}

\bibitem{Cong2020}
Jian Cong, Shan Yang, Lei Xie, et~al.,
\newblock ``{Data Efficient Voice Cloning from Noisy Samples with Domain Adversarial Training},''
\newblock in {\em Proceedings of {ISCA} {INTERSPEECH}}, Shanghai, China, 2020, pp. 811--815.

\bibitem{Ren2021}
Yi~Ren, Jinglin Liu, and Zhou Zhao,
\newblock ``{Portaspeech: Portable and High-Quality Generative Text-to-Speech},''
\newblock in {\em Proceedings of {MIT} Press NeurIPS}, Virtual Event, 2021, pp. 13963--13974.

\bibitem{Casanova2021}
Edresson Casanova, Christopher Shulby, Eren G{\"{o}}lge, et~al.,
\newblock ``{Sc-Glowtts: an Efficient Zero-Shot Multi-Speaker Text-to-Speech Model},''
\newblock in {\em Proceedings of {ISCA} {INTERSPEECH}}, Brno, Czechia, 2021, pp. 3645--3649.

\bibitem{Wu2022}
Yihan Wu, Xu~Tan, Bohan Li, et~al.,
\newblock ``{Adaspeech 4: Adaptive Text to Speech in Zero-Shot Scenarios},''
\newblock in {\em Proceedings of {ISCA} {INTERSPEECH}}, Incheon, Korea, 2022, pp. 2568--2572.

\bibitem{Casanova2022}
Edresson Casanova, Julian Weber, Christopher~Dane Shulby, et~al.,
\newblock ``{Yourtts: Towards Zero-Shot Multi-Speaker Tts and Zero-Shot Voice Conversion for Everyone},''
\newblock in {\em Proceedings of {ACM} {ICML}}, Baltimore, Maryland, USA, 2022, pp. 2709--2720.

\bibitem{Li2023}
Yinghao~Aaron Li, Cong Han, Vinay~S. Raghavan, et~al.,
\newblock ``{Styletts 2: Towards Human-Level Text-to-Speech through Style Diffusion and Adversarial Training with Large Speech Language Models},''
\newblock in {\em Proceedings of {MIT} Press NeurIPS}, New Orleans, LA, USA, 2023.

\bibitem{Ju2024}
Zeqian Ju, Yuancheng Wang, Kai Shen, et~al.,
\newblock ``{Naturalspeech 3: Zero-Shot Speech Synthesis with Factorized Codec and Diffusion Models},''
\newblock 2024.

\bibitem{Jiang2024}
Ziyue Jiang, Jinglin Liu, Yi~Ren, et~al.,
\newblock ``{Mega-Tts 2: Boosting Prompting Mechanisms for Zero-Shot Speech Synthesis},''
\newblock in {\em Proceedings of {ICLR}}, Vienna, Austria, 2024.

\bibitem{2022Huang}
Rongjie Huang, Yi~Ren, Jinglin Liu, Chenye Cui, and Zhou Zhao,
\newblock ``{Generspeech: Towards Style Transfer for Generalizable Out-of-Domain Text-to-Speech},''
\newblock in {\em Proceedings of {MIT} Press NeurIPS}, New Orleans, LA, USA [hybrid], 2022.

\bibitem{lee2023}
Sang{-}Hoon Lee, Ha{-}Yeong Choi, Seung{-}Bin Kim, and Seong{-}Whan Lee,
\newblock ``{Hierspeech++: Bridging the Gap Between Semantic and Acoustic Representation of Speech by Hierarchical Variational Inference for Zero-Shot Speech Synthesis},''
\newblock {\em CoRR}, vol. abs/2311.12454, 2023.

\bibitem{2023Kang}
Minki Kang, Dongchan Min, and Sung~Ju Hwang,
\newblock ``{Grad-Stylespeech: Any-Speaker Adaptive Text-to-Speech Synthesis with Diffusion Models},''
\newblock in {\em Proceedings of {IEEE} {ICASSP}}, Rhodes Island, Greece, 2023, pp. 1--5.

\bibitem{Shen2024}
Kai Shen, Zeqian Ju, Xu~Tan, et~al.,
\newblock ``{Naturalspeech 2: Latent Diffusion Models Are Natural and Zero-Shot Speech and Singing Synthesizers},''
\newblock in {\em Proceedings of {ICLR}}, Vienna, Austria, 2024.

\bibitem{2024Lee}
Keon Lee, Dong~Won Kim, Jaehyeon Kim, and Jaewoong Cho,
\newblock ``{Ditto-Tts: Efficient and Scalable Zero-Shot Text-to-Speech with Diffusion Transformer},''
\newblock {\em CoRR}, vol. abs/2406.11427, 2024.

\bibitem{Wang2023}
Chengyi Wang, Sanyuan Chen, Yu~Wu, et~al.,
\newblock ``Neural codec language models are zero-shot text to speech synthesizers,''
\newblock {\em CoRR}, vol. abs/2301.02111, 2023.

\bibitem{Chen2024}
Sanyuan Chen, Shujie Liu, Long Zhou, et~al.,
\newblock ``{Vall-E 2: Neural Codec Language Models Are Human Parity Zero-Shot Text to Speech Synthesizers},''
\newblock {\em CoRR}, vol. abs/2406.05370, 2024.

\bibitem{CosyVoice}
Zhihao Du, Qian Chen, Shiliang Zhang, et~al.,
\newblock ``{Cosyvoice: a Scalable Multilingual Zero-Shot Text-to-Speech Synthesizer Based on Supervised Semantic Tokens},''
\newblock {\em CoRR}, vol. abs/2407.05407, 2024.

\bibitem{GPT-SoVITS}
RVC-Boss,
\newblock ``Gpt-sovits,'' https://github.com/RVC-Boss/GPT-SoVITS/tree/b3e8eb40c25bea1b9977195a380d40c941040419, 2024,
\newblock Accessed: 2024-08-15.

\bibitem{seedtts}
Philip Anastassiou, Jiawei Chen, Jitong Chen, et~al.,
\newblock ``{Seed-Tts: a Family of High-Quality Versatile Speech Generation Models},''
\newblock {\em CoRR}, vol. abs/2406.02430, 2024.

\bibitem{Jia2018}
Ye~Jia, Yu~Zhang, Ron~J. Weiss, et~al.,
\newblock ``{Transfer Learning from Speaker Verification to Multispeaker Text-to-Speech Synthesis},''
\newblock in {\em Proceedings of {MIT} Press NeurIPS}, Montréal, Canada, 2018, pp. 4485--4495.

\bibitem{Arik2018}
Sercan~{\"{O}}mer Arik, Jitong Chen, Kainan Peng, et~al.,
\newblock ``{Neural Voice Cloning with a Few Samples},''
\newblock in {\em Proceedings of {MIT} Press NeurIPS}, Montréal, Canada, 2018, pp. 10040--10050.

\bibitem{Kumar2022}
Neeraj Kumar, Ankur Narang, and Brejesh Lall,
\newblock ``{Zero-Shot Normalization Driven Multi-Speaker Text to Speech Synthesis},''
\newblock {\em {IEEE} {ACM} Trans. Audio Speech Lang. Process.}, vol. 30, pp. 1679--1693, 2022.

\bibitem{Wang17}
Yuxuan Wang, R.~J. Skerry{-}Ryan, Daisy Stanton, et~al.,
\newblock ``{Tacotron: Towards End-to-End Speech Synthesis},''
\newblock in {\em Proceedings of {ISCA} {INTERSPEECH}}, Stockholm, Sweden, 2017, pp. 4006--4010.

\bibitem{Shaw18}
Peter Shaw, Jakob Uszkoreit, and Ashish Vaswani,
\newblock ``{Self-Attention with Relative Position Representations},''
\newblock in {\em Proceedings of Association for Computational Linguistics {NAACL}}, New Orleans, Louisiana, USA, 2018, pp. 464--468.

\bibitem{chen2024eres2netv2}
Yafeng Chen, Siqi Zheng, Hui Wang, et~al.,
\newblock ``Eres2netv2: Boosting short-duration speaker verification performance with computational efficiency,''
\newblock 2024.

\bibitem{zheng20233d}
Siqi Zheng, Luyao Cheng, Yafeng Chen, et~al.,
\newblock ``{3d-Speaker: a Large-Scale Multi-Device, Multi-Distance, and Multi-Dialect Corpus for Speech Representation Disentanglement},''
\newblock 2023, vol. abs/2306.15354.

\bibitem{Chen2020}
Ting Chen, Simon Kornblith, Mohammad Norouzi, et~al.,
\newblock ``{a Simple Framework for Contrastive Learning of Visual Representations},''
\newblock in {\em Proceedings of {ACM} {ICML}}, Virtual Event, 2020, pp. 1597--1607.

\bibitem{wangNSFall}
Xin Wang, Shinji Takaki, and Junichi Yamagishi,
\newblock ``{Neural Source-Filter Waveform Models for Statistical Parametric Speech Synthesis},''
\newblock {\em IEEE/ACM TASLP}, vol. 28, pp. 402--415, 2020.

\bibitem{xtts}
Coqui,
\newblock ``Xtts-v2,'' https://huggingface.co/coqui/XTTS-v2,
\newblock Accessed: 2024-08-26.

\bibitem{ecapaTDNN}
Brecht Desplanques, Jenthe Thienpondt, and Kris Demuynck,
\newblock ``{Ecapa-Tdnn: Emphasized Channel Attention, Propagation and Aggregation in Tdnn Based Speaker Verification},''
\newblock in {\em Proceedings of {ISCA} {INTERSPEECH}}, Shanghai, China, 2020, pp. 3830--3834.

\bibitem{gao2022paraformer}
``{Paraformer: Fast and Accurate Parallel Transformer for Non-Autoregressive End-to-End Speech Recognition},''
\newblock in {\em Proceedings of {ISCA} {INTERSPEECH}}, Incheon, Korea, 2022, pp. 2063--2067.

\end{thebibliography}
